\documentclass{DISproc}

\newcommand{\abr}[1]{\langle #1\rangle}

\newcommand{\sbr}[1]{\left[ #1\right]}
\newcommand{\done}{{\rm d}}
\newcommand{\order}{\mathcal{O}}
\newcommand{\mc}[1]{\mathcal{#1}}

\newcommand{\MCatNLO}{M\protect\scalebox{0.8}{C}@N\protect\scalebox{0.8}{LO}\xspace}

\newcommand{\BlackHat}{B\protect\scalebox{0.8}{LACK}H\protect\scalebox{0.8}{AT}\xspace}

\newcommand{\Sherpa}{S\protect\scalebox{0.8}{HERPA}\xspace}

\newcommand{\ATLAS}{ATLAS\xspace}

\long\def\symbolfootnote[#1]#2{\begingroup%
\def\thefootnote{\fnsymbol{footnote}}\footnote[#1]{#2}\endgroup}

\newcommand{\im}{\imath}
\newcommand{\jm}{\jmath}
\newcommand{\args}[1]{\vec{#1}}
\newcommand{\argc}[2]{\vec{#1}_{\rm #2}}

\newcommand{\bmap}[3]{b_{#1}(#3)}
\newcommand{\rmap}[3]{r_{\widetilde{#1}}(#3)}

\newcommand{\nnb}{\nonumber}
\newcommand{\bea}{\begin{eqnarray}}
\newcommand{\eea}{\end{eqnarray}}
\newcommand{\bi}{\begin{itemize}}
\newcommand{\ei}{\end{itemize}}

\begin{document}

\title{\hfill\begin{small}SLAC-PUB-15097\end{small}\\
W+n-jet predictions at NLO matched with a parton shower
  }

\author{{\slshape Frank Siegert$^1$, Stefan H\"oche$^2$, Frank Krauss$^3$, Marek Sch\"onherr$^3$}\\[1ex]
$^1$Physikalisches Institut, Albert-Ludwigs-Universit{\"a}t Freiburg, D-79104 Freiburg, Germany\\
$^2$SLAC National Accelerator Laboratory, Menlo Park, CA 94025, USA\\
$^3$Institute for Particle Physics Phenomenology, Durham University, Durham DH1 3LE, UK }

\contribID{242}

\doi  

\maketitle

\begin{abstract}
  The MC@NLO method as implemented in the Sherpa MC generator is
  presented using the production of W-bosons in conjunction with up to
  three jets as an example. Corresponding results computed at next-to
  leading order in QCD and including parton shower corrections are
  compared to recent experimental data from the Large Hadron Collider.
\end{abstract}

\section{Introduction}

To make the LHC a discovery machine we have to acknowledge the fact that it is
a QCD machine. Many signals suffer from large backgrounds largely due to QCD
multi-jet production which have to be under good theoretical control to
interpret the measurements.

There are mainly two approaches to include higher-order QCD corrections in
theoretical calculations of scattering matrix elements (ME):

{\bf Fixed-order ME calculations} put an emphasis on the
exact evaluation of a finite number of terms in the perturbation series.
Apart from being exact to the given order this also
has the benefit of including all interference terms from different diagrams and
working with a number of colours $N_C=3$. Last, but not least, it becomes
possible to include also the exact finite part of virtual contributions in a
fixed-order calculation.

Their disadvantages appear when an observable becomes sensitive to
logarithmically enhanced regions. It is not sufficient to truncate the
perturbation series at a fixed order anymore, if the perturbative nature
of the coupling constant $\alpha_s$ is countered by large logarithms which
appear when partons become soft or collinear to each other. This problem is
solved in the {\bf parton shower} approach (PS), where the logarithmically enhanced
contributions are resummed to all orders, albeit only in an approximation
valid in the collinear limit of the matrix element and in the large $N_C$
limit. This allows to generate events with partons at the hadronisation scale
and thus enables exclusive hadron-level event generation.

It is thus a worthwhile goal to combine the two approaches and keep the
advantages: Include the virtual contributions and first hard emission from the
exact next-to-leading order matrix element, and add further parton evolution
using a parton shower approach.

\section{Recap: Resummation and NLO calculations}

The basic property of QCD allowing a parton shower resummation is the universal factorisation of real emission matrix elements in the collinear limit:
\begin{equation}
  \mathcal{R} \;\mathop{\longrightarrow}^{ij\text{ collinear}}\; \mathcal{D}_{ij}^{\rm(PS)}=\mathcal{B} \times \left(
    \frac{1}{2p_i p_j} \;
    8\pi\alpha_s \;
    \mathcal{K}_{ij}(p_i,p_j) \right) 
\end{equation}

With the approximation that multiple emissions happen independently of each other (thus yielding Poisson statistics) the corresponding branching probability can be exponentiated to give the total no-branching probability down to an evolution scale $t$:
\begin{align}
  \Delta^{\rm(PS)}(t)\,&=\;\prod_{\widetilde{\im\jm}}\,\exp\left\{-\sum_{f_i=q,g}
    \int\done\Phi_{R|B}^{ij}\,\Theta\left(t(\Phi_{R|B}^{ij})-t\right)\,
    \frac{\mathcal{D}_{ij}^{\rm(PS)}}{\mathcal{B}}\right\}
\end{align}

To understand the implications of the no-branching probability $\Delta$, let
us look at the expectation value of an observable $O$ taking into
account up to one emission from the parton shower on top of a Born-level event:
\begin{align}
  \small
  \label{eq:ops}
  \abr{O}^{\rm (PS)}\,=\,\int\done\Phi_B\,
  \mathcal{B}\Bigg[\underbrace{
    \Delta^{\rm(PS)}(t_0)\,O(\Phi_B)}_{\text{unresolved}}\,
  \;+\;
  \underbrace{
    \sum_{\widetilde{\im\jm}}
    \sum_{f_i}
    \int_{t_0}^{\mu_F^2}\done\Phi_{R|B}^{ij}\,
    \frac{\mathcal{D}_{ij}^{\rm(PS)}}{\mathcal{B}}
    \Delta^{\rm(PS)}(t)\,
    O\left(r_{\widetilde{\im\jm}}(\Phi_B)\right)
  }_{\text{resolved}}
  \Bigg]
\end{align}
The ``unresolved'' contribution corresponds to the event generation case where
no emission above the parton shower cut-off scale $t_0$ has been generated and
is thus proportional to the no-branching probability $\Delta(t_0)$. The
``resolved'' contribution on the other hand represents the integration over
events which had an emission with evolution scale $t>t_0$.

As a reminder and to fix some notation, the contributions of an NLO calculation
for the expectation value of $O$ are briefly recalled:
\begin{equation}
  \small
  \label{eq:onlo}
  \begin{split}
  \abr{O}^{\rm (NLO)}\,=&\;\sum_{\argc{f}{B}}\int\done\Phi_B\,
  \sbr{\,\mathcal{B}(\Phi_B)+\tilde{\mathcal{V}}(\Phi_B)+
    \sum_{\widetilde{\im\jm}}\mathcal{I}_{\widetilde{\im\jm}}^{\rm(S)}(\Phi_B)\,}\,O(\Phi_B)\\
  &\qquad+\sum_{\argc{f}{R}}\int\done\Phi_R\,
  \sbr{\,\mathcal{R}(\Phi_R)\,O(\Phi_R)-\sum_{\{ij\}}\mathcal{D}_{ij}^{\rm(S)}(\Phi_R)\,O(\bmap{ij}{k}{\Phi_R})\,}
  \end{split}
\end{equation}

Here, the Born ($\mathcal{B}$), virtual ($\mathcal{V}$) and real emission
($\mathcal{R}$) pieces are denoted accordingly. Since $\mathcal{V}$ and
$\mathcal{R}$ are oppositely divergent in separate phase space integrations,
one needs to employ a subtraction procedure: The real subtraction terms $\mathcal{D}$
are linked to their integrated form $\mathcal{I}$ by a phase space integration
over the 1-emission phase space and can be calculated e.g.~in the scheme of~\cite{Catani:1996vz}.

\section{Combining NLO and PS}

Applying PS resummation on top of such NLO events is not straightforward: The
observable has to be evaluated at different kinematics in the $\mc{R}$
and $\mc{D}$ terms. But if that is taken into account, and they are thus
showered separately, one generates an additional term at $\order(\alpha_s)$~\cite{Frixione:2002ik}
which is not present in the NLO calculation (``double counting''). To counter
that term a solution was proposed in~\cite{Frixione:2002ik} introducing an additional set of ``modified'' subtraction terms $\mc{D}^{\rm(A)}$. When generating events according to that modified NLO cross section, they will have either $\Phi_R$ kinematics (resolved, non-singular term) and are kept as they are or $\Phi_B$ kinematics. In the latter case, they are processed through a one-step PS with $\Delta^{\rm(A)}$, i.e.~using the modified subtraction terms as PS kernels, either generating an emission (resolved, singular) or no emission (unresolved, singular) above $t_0$. The result of this procedure,
\begin{equation}
  \small
  \begin{split}
  \abr{O}^{\rm(NLO+PS)}=&\sum_{\args{f}_B}\int\done\Phi_B
  \bar{\mc{B}}^{\rm(A)}(\Phi_B)\left[\vphantom{\Bigg)_{\int}^{\int}}\right.
    \underbrace{\Delta^{\rm(A)}(t_0)}_{\text{unresolved}}\,
    O(\Phi_B)\nnb
  +\sum_{\{\widetilde{\im\jm},f_i\}}
    \int\limits_{t_0}\done\Phi_{R|B}^{ij}
    \underbrace{
    \frac{\mc{D}_{ij}^{\rm(A)}(\rmap{\im\jm}{k}{\Phi_{B}})}{
      \mc{B}(\Phi_B)}
    \Delta^{\rm(A)}(t)}_{\text{resolved, singular}}
    O(\rmap{\im\jm}{k}{\Phi_B})
    \left.\vphantom{\Bigg)_{\int}^{\int}}\right]\nnb\\
  &\quad+\;\sum_{\args{f}_R}\int\done\Phi_R\,
    \underbrace{\sbr{\,\mc{R}(\Phi_R)-\sum_{ij}\mc{D}^{\rm(A)}_{ij}(\Phi_R)\,}}_{\text{resolved, non-singular}}\,
    O(\Phi_R),
  \end{split}
\end{equation}
with $\bar{\mc{B}}^{\rm(A)}(\Phi_B)$ defined as
\begin{equation*}
  \small
    \bar{\mc{B}}^{\rm(A)}(\Phi_B)\;=\,\mc{B}(\Phi_B)+\tilde{\mc{V}}(\Phi_B)+
      \sum_{\{\widetilde{\im\jm}\}}\mc{I}_{\widetilde{\im\jm}}^{\rm(S)}(\Phi_B)
    +\sum_{\{\widetilde{\im\jm}\}}
      \sum_{f_i=q,g}\int\done\Phi_{R|B}^{ij}\;
      \sbr{\,\mc{D}^{\rm(A)}_{ij}(\rmap{\im\jm}{k}{\Phi_B})
        -\mc{D}^{\rm(S)}_{ij}(\rmap{\im\jm}{k}{\Phi_B})\,}
\end{equation*}
can be shown to reproduce $\abr{O}^{\rm(NLO)}$ to $\order(\alpha_s)$.

This procedure still leaves the freedom of choosing $\Delta^{\rm(A)}$. The original approach~\cite{Frixione:2002ik} uses the parton shower splitting kernels as additional subtraction terms, $\mc{D}^{\rm(A)}_{ij}\to\mc{D}^{\rm(PS)}_{ij}$. This option has the advantage that the exponentiation in the ``resolved, singular'' contribution is naturally bounded by the factorisation scale $\mu_F$. Problems appear though with soft divergences in the ``resolved, non-singular'' integration, since the parton shower splitting kernels do not cover soft divergences.

An alternative approach was suggested in~\cite{Hoeche:2011fd} and implemented in \Sherpa~\cite{Gleisberg:2008ta}, where the full Catani-Seymour dipoles are used $\mc{D}^{\rm(A)}_{ij}\to\mc{D}^{\rm(S)}_{ij}$. With this, $\bar{\mc{B}}^{\rm(A)}$ simplifies significantly, but at a cost: $\mc{D}^{\rm(S)}$ can become negative, resulting in $\Delta>1$. This is generated in \Sherpa by a weighted $N_C=3$ one-step PS based on the subtraction terms $\mc{D}^{\rm(S)}$. With this approach, exact NLO accuracy also for sub-leading colour configurations is achieved. The phase space boundary for the exponentiation though has to be imposed ``artificially'' by cuts in the dipole phase space.

\section{Results}

Results for $W+n$-jet production at the LHC are presented here in comparison to \ATLAS data~\cite{Aad:2012en}. Events are simulated using \Sherpa's \MCatNLO for $W+0$, $W+1$, $W+2$ and $W+3$-jet production with virtual corrections from \BlackHat~\cite{Berger:2008sj} including a leading-colour approximation for the $W+3$-jet virtual ME. For $n>0$ events are generated requiring $k_T$ jets with $p_\perp>10$~GeV, and the exponentiation region was restricted using an $\alpha=0.01$-cut in the dipole terms~\cite{Nagy:2003tz}. The comparison comprises of three levels of event simulation: ``NLO'' as fixed-order calculation, ``MC@NLO 1em'' as \MCatNLO including the hardest emission only and ``MC@NLO PL'' as \MCatNLO including the full PS. All distributions are simulated at NLO accuracy and the theoretical predictions describe data very well.


\begin{minipage}{0.49\linewidth}
  {
    \centering
    \includegraphics[width=0.8\linewidth]{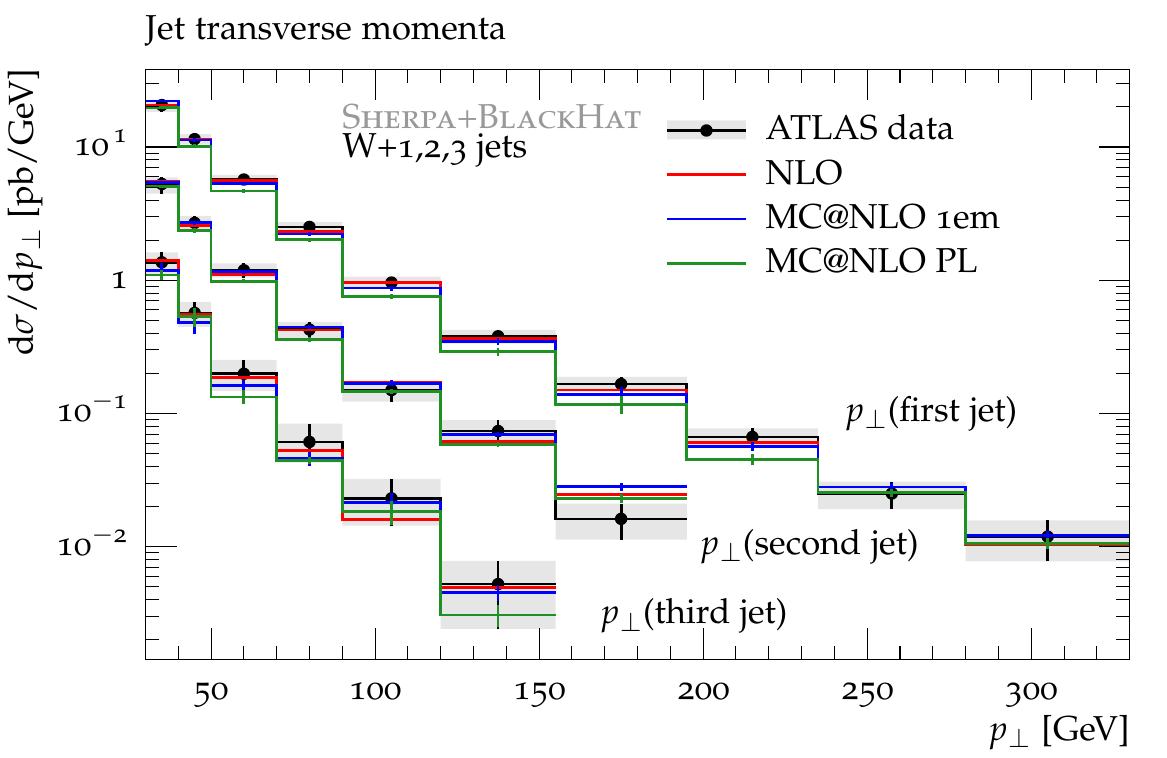}\vspace*{-3mm}\\
    \includegraphics[width=0.8\linewidth]{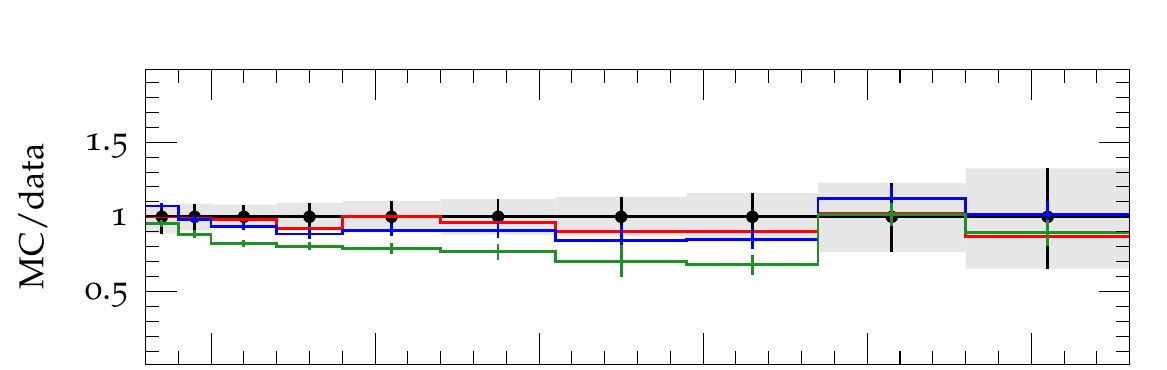}\\
    \includegraphics[width=0.8\linewidth]{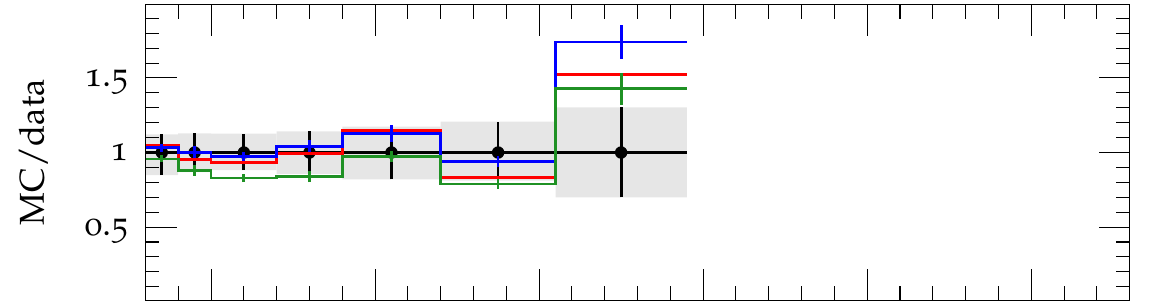}\\
    \includegraphics[width=0.8\linewidth]{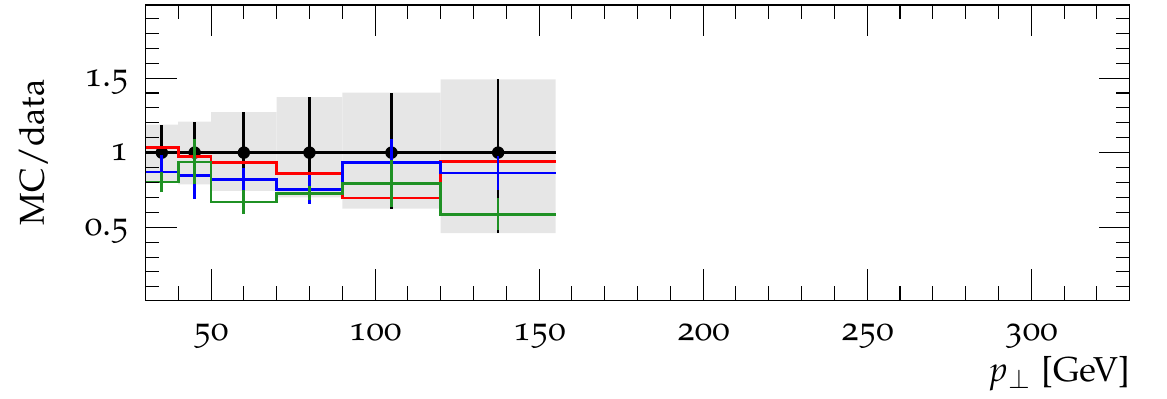}\\
  }
    Fig.~1: Transverse momentum of the first, second and third jet (from top
    to bottom).
\end{minipage}\nolinebreak
\begin{minipage}{0.02\linewidth}
  \hfill
\end{minipage}\nolinebreak
\begin{minipage}{0.49\linewidth}
  {
    \centering
    \includegraphics[width=0.7\linewidth]{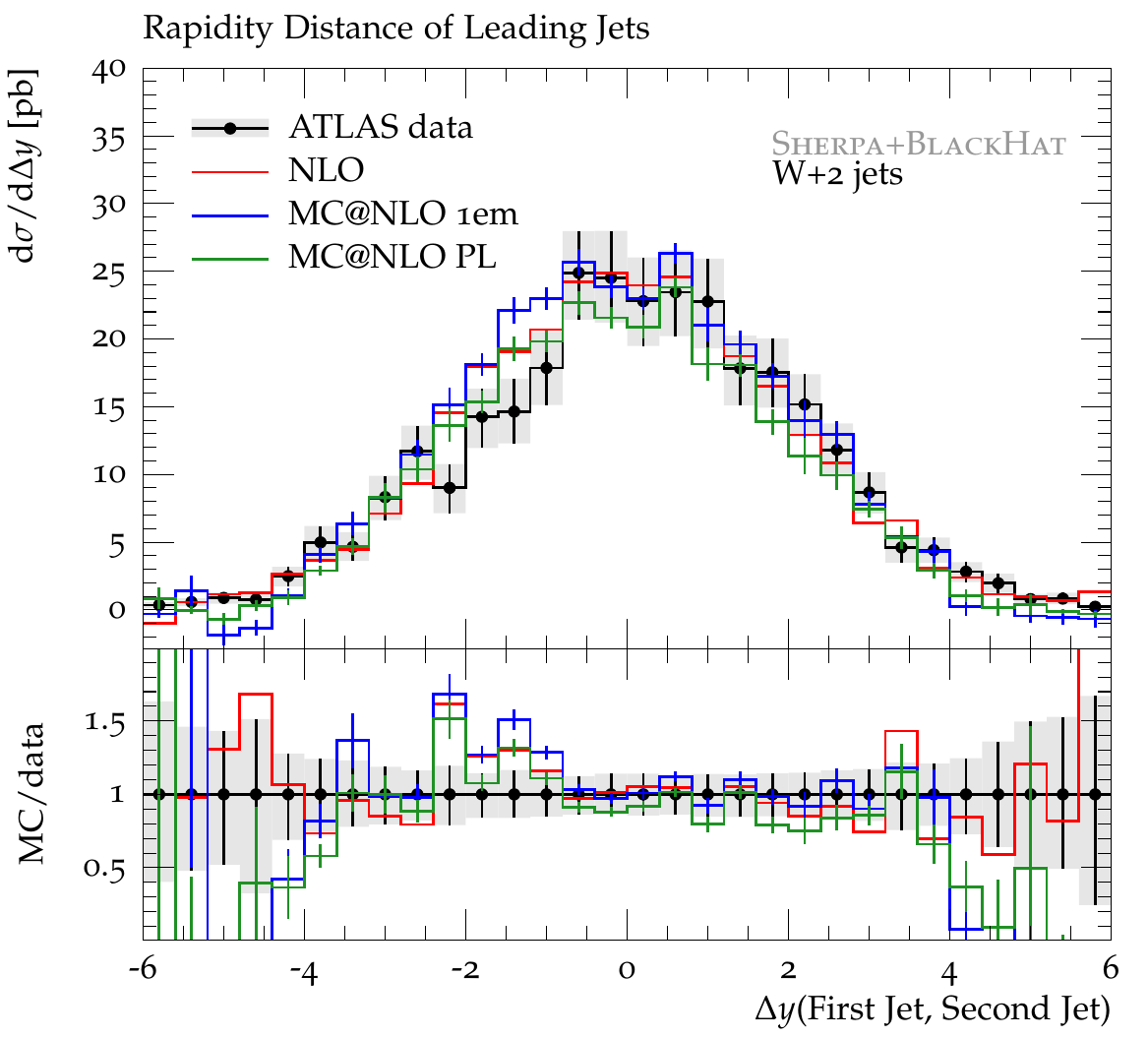}\vspace*{-3mm}\\
    \includegraphics[width=0.7\linewidth]{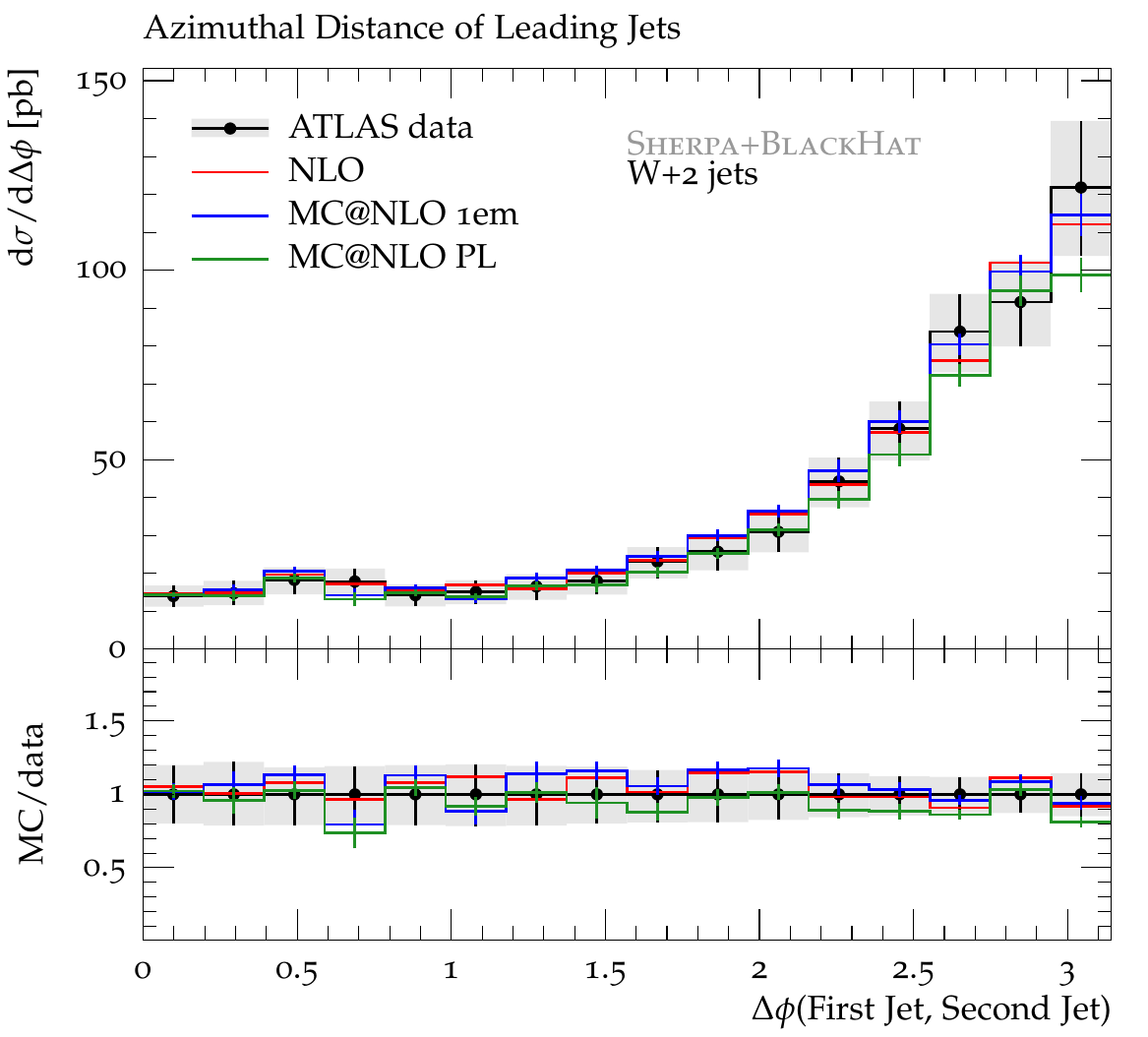}\\
  }
  Fig.~2: Angular correlations of the two leading jets in
  $W^\pm+\geq 2$ jet production.
\end{minipage}

\section{Conclusions and outlook}

\vspace*{-2mm}
The concept of NLO+PS matching was presented in a common formalism, and  \MCatNLO was developed as a special case. It was emphasised that an exact treatment of sub-leading colour configurations can be achieved by exponentiating Catani-Seymour subtraction terms. The first NLO+PS predictions for $W$+3 jets were presented and showed good agreement with experimental data from \ATLAS. With this method at hand, it becomes now possible to generate exclusive hadron-level event samples at NLO accuracy.

In the future, an improved functional form of the phase-space cut $\alpha$ will allow for a better limitation of the exponentiation region. Also, by merging NLO+PS with higher-multiplicity tree-level MEs one can already provide an improved description of multi-jet final states in the MENLOPS approach~\cite{Hoeche:2010kg}. Ultimately, it remains a goal to achieve the merging of NLO+PS predictions at different multiplicities into one inclusive sample.

\vspace*{-2mm}
{\raggedright
\begin{footnotesize}
\bibliographystyle{DISproc}
\bibliography{siegert_frank.bib}
\end{footnotesize}
}

\end{document}